\documentclass[twocolumn,showpacs,preprintnumbers,amsmath,amssymb]{revtex4}

\usepackage{graphicx}
\usepackage{dcolumn}
\usepackage{bm}
\usepackage{epsfig}
\newcommand{\bea}{\begin{eqnarray}}
\newcommand{\eea}{\end{eqnarray}}
\begin{document}

\preprint{APS/123-QED}

\title{Some Exact Solutions to
Equations of Motion\\ of an Incompressible Third Grade Fluid}

\author{Saifullah}
 \email{saifullahkhalid75@yahoo.com}
\affiliation{%
School of Mathematical Sciences\\ Government College
University\\ Lahore, Pakistan.}%


\date{\today}

\begin{abstract}
This investigation deals with some exact solutions of the equations
governing the steady plane motions of an incompressible third grade
fluid by using complex variables and complex functions. Some of the
solutions admit, as particular cases, all the solutions of Moro et
al.
\end{abstract}

\maketitle

\section{\label{sec:level1}Introduction}

The governing equations that describe flows of Newtonian fluids are
the Navier-Stokes equations. The mechanical behavior of many real
fluids, especially those of low molecular weight, appears to be
accurately described by these equations over a wide range of
circumstances. There are, however, many real substances which are
capable of flowing but which are not at all well described by the
Navier-Stokes theory. Due to this reason many fluid models have been
proposed and studied by different authors. Among these, the fluids
of differential type have received much attention. Interesting
studies of differential type fluids are given by Rajagopal
\cite{Rajagopal1}, Erdogan \cite{Erdogan}, Rajagopal and Gupta
\cite{Gupta}, Bandelli \cite{Bandelli1}, Siddiqui and Kaloni
\cite{Siddiqui0}, Benharbit and Siddiqui \cite{Siddiqui1}, Ariel
\cite{Ariel0, Ariel1}, Fetecau and Fetecau \cite{Fetecau} and Hayat
et al. \cite{Hayat3, Hayat4, Hayat5}. The fluids of third grade,
which form a subclass of the fluids of differential type, have been
successfully studied in various types of motions.

Moro et al. \cite{Moro} determined some exact solutions of the
equations governing the steady plane motion of an incompressible
third grade fluid employing hodograph and Legendre transformations.

The aim of this paper is to present some exact solutions for steady
plane motions of incompressible third grade fluids. For this, the
complex variables and complex functions are used. Some previous
solutions can be obtained as special cases of our solutions. In
sections 2 and 3 are presented the constitutive equations and the
expression of the vorticity function $\omega.$ Section 4 contains
solutions and their graphical illustrations.

\section{\label{sec:level2}Constitutive and governing equations}

The fluids of grade $n,$ introduced by Rivlin and Ericksen
\cite{Rivlin}, are the fluids for which the stress tensor is a
polynomial of degree $n$ in the first $n$ Rivlin-Ericksen tensors
defined recursively by \bea \nonumber &&A_1 = (\partial_i u_j
+\partial_j u_i)_{i,\, j}\,;\\&& A_n = \frac{d}{dt}A_{n-1} + A_{n-1}
L + L^t A_{n-1}\,, \,\,\, n > 1 \eea where $\frac{d}{dt}=\partial_t + u
\cdot\nabla$ denotes the material derivative and
\begin{equation}
L=(\partial_j u_i)_{i,\,j} \, \, ; \,\,\,\,\,\,\, L^t=(\partial_i
u_j)_{i,\,j} \, .\nonumber
\end{equation}
Physical considerations were taken into account by Fosdick and
Rajagopal \cite{Fosdic} in order to obtain the following form of
constitutive equation for the Cauchy stress $T$ in a third grade
fluid: \bea\nonumber &&T = -pI + \mu
A_1+\alpha_1A_2+\alpha_2A^2_1+\beta_1 A_3+\\&&+
\beta_2[A_1A_2+A_2A_1]+\beta_3(tr A^2_1)A_1 \eea where $-pI$ is the
spherical stress due to the constraint of incompressibility, $\mu$
is the dynamic viscosity, $\alpha_i(i=1,2)$ and $\beta_i(i=1,2,3)$
are the material constants and the three
Rivlin-Ericksen tensors $A_1, A_2$ and $A_3$ are given in equation (1).\\
Furthermore, a complete thermodynamic analysis of the constitutive
equation (2) has been given by Fosdick and Rajagopal \cite{Fosdic}.
The Clausius--Duhem inequality and the assumption that the Helmholtz
free energy is a minimum in equilibrium provide the following
restrictions, \bea\nonumber &&\mu \geq 0\, ,\,\,\,\alpha_1 \geq 0\,
,\,\,\,|\alpha_1 + \alpha_2| \leq \sqrt{24 \mu \beta_3}\,,\\&&
\beta_1=\beta_2=0\, ,\,\,\,\beta_3 \geq 0\,. \eea Thus equation (2)
becomes
\begin{equation}
T=-pI+\alpha_1 A_2 + \alpha_2 A^2_1 +[\mu +\beta_3(tr A^2_1)]A_1\, ,
\nonumber
\end{equation}
where $\mu_{eff}= \mu + \beta_3(trA^2_1)$ is the effective
shear-dependent viscosity.\\
The velocity field corresponding to the motion is given as:
$$v=v(x,y)=(u(x,y), v(x,y), 0)\,.$$

\section{\label{sec:level3}Flow equations}

The basic equations governing the steady plane motion of a
homogeneous incompressible fluid of third grade, in the absence of
body forces are \cite{Moro} \bea \frac{\partial u}{\partial x} +
\frac{\partial v}{\partial y}= 0 \eea \bea\nonumber&& \frac{\partial
h}{\partial x}=\rho v \omega - \mu\frac{\partial \omega}{\partial
y}- \alpha_{1}v \nabla^{2}\omega - \beta_{3}\frac{\partial (\omega
M)}{\partial y}+\\&&+ 2\beta_{3}(\frac{\partial u}{\partial
x}\frac{\partial M}{\partial x}+\frac{\partial v}{\partial
x}\frac{\partial M}{\partial y}) \eea \bea\nonumber&& \frac{\partial
h}{\partial y}=-\rho u \omega+\mu\frac{\partial \omega}{\partial
x}+\alpha_{1}u \nabla^{2}\omega+\beta_{3}\frac{\partial (\omega
M)}{\partial x}+\\&&+ 2\beta_{3}(\frac{\partial u}{\partial
y}\frac{\partial M}{\partial x}+\frac{\partial v}{\partial
y}\frac{\partial M}{\partial y}) \eea where
\begin{equation}
\omega = \frac{\partial v}{\partial x}-\frac{\partial u}{\partial
y}\nonumber
\end{equation}
\begin{equation}
h=\frac{\rho q^{2}}{2}-\alpha_{1}(u \nabla^{2}u+v\nabla^{2}v)-(3
\alpha_{1}+2\alpha_{2})\frac{M}{4}+p
\end{equation}
\begin{equation}
M=4(\frac{\partial u}{\partial x})^{2}+4(\frac{\partial v}{\partial
y})^{2}+2(\frac{\partial v}{\partial x} +\frac{\partial u}{\partial
y})^{2}\nonumber
\end{equation}
\begin{equation}
q^{2}=u^{2}+v^{2}\nonumber
\end{equation}
Equations (5) and (6) are non-linear partial differential equations
for three unknowns $u,\,v$   and $\rho$  as functions of $x$ and
$y$. In equations (5), (6) and (7), the viscosity $\mu$ and the
material constants
$\alpha_{1},\,\alpha_{2},\,\beta_{1},\,\beta_{2},\,\beta_{3}$
satisfy the constraints given in equation (3) \cite{Fosdic, Truessdell}.\\
Equation (4) implies the existence of a stream function $\psi(x,y)$
such that
\begin{equation}
u=\frac{\partial \psi}{\partial y}\,\,\,\mbox{and}\,\,\,
v=-\frac{\partial \psi}{\partial x}
\end{equation}
Equations (4) and (5), on utilizing equation (8) and the
compatibility condition $ \frac{\partial^{2}h}{\partial x
\partial y}= \frac{\partial^{2}h}{\partial y \partial x}\,\, ,
$ yield \bea\nonumber &&\rho\Big(\frac{\partial \psi}{\partial
y}\frac{\partial \omega}{\partial x} -\frac{\partial \psi}{\partial
x}\frac{\partial \omega}{\partial
y}\Big)-\alpha_{1}\Big\{\frac{\partial \psi}{\partial
y}\frac{\partial (\nabla^{2}\omega)}{\partial x}-\frac{\partial
\psi}{\partial x}\frac{\partial (\nabla^{2}\omega)}{\partial
y}\Big\}\\&&\nonumber -\beta_{3}\Big\{\frac{\partial^{2}(\omega
M)}{\partial x^{2}}+\frac{\partial^{2}(\omega M)}{\partial
y^{2}}\Big\}+2\beta_{3}\Big\{2\frac{\partial^{2}\psi}{\partial
x\partial y}\frac{\partial^{2}M}{\partial y\partial x}-\\&&
-\frac{\partial^{2}\psi}{\partial
x^{2}}\frac{\partial^{2}M}{\partial
y^{2}}-\frac{\partial^{2}\psi}{\partial
y^{2}}\frac{\partial^{2}M}{\partial
x^{2}}\Big\}-\mu\nabla^{2}\omega= 0 \eea where
\begin{equation}
\omega=-\Big(\frac{\partial^{2}\psi}{\partial
x^{2}}+\frac{\partial^{2}\psi}{\partial y^{2}}\Big)\nonumber
\end{equation}
\begin{equation}
M=8\frac{\partial^{2} \psi^{2}}{\partial y\partial
x}+2\Big(\frac{\partial^{2}\psi}{\partial
y^{2}}-\frac{\partial^{2}\psi}{\partial x^{2}}\Big)\nonumber
\end{equation}
Let
\begin{equation}
z=x+\iota y
\,\,\,\,\,\,\,\,\,\,\,\,\,\,\,\mbox{and}\,\,\,\,\,\,\,\,\,\,\,\,\,\,\,
\overline{z}=x-\iota y\nonumber
\end{equation}
Then from Stallybrass \cite{Stallybrass} \bea && \nonumber
2\frac{\partial(\bullet)}{\partial\overline{z}}=\frac{\partial(\bullet)}{\partial
x}+\iota\frac{\partial(\bullet)}{\partial
y},\,\,\,\,\,\,\,\,\,\,2\frac{\partial(\bullet)}{\partial
z}=\frac{\partial(\bullet)}{\partial
x}-\iota\frac{\partial(\bullet)}{\partial
y},\\&&4Im\Big\{\frac{\partial(\bullet)}{\partial
\overline{z}}\frac{\partial(\circ)}{\partial
z}\Big\}=\frac{\partial(\bullet)}{\partial
y}\frac{\partial(\circ)}{\partial
x}-\frac{\partial(\bullet)}{\partial
x}\frac{\partial(\circ)}{\partial y} \eea Equation (9), on utilizing
equation (10), becomes \bea\nonumber&& Im\Big\{\frac{\partial
\psi}{\partial \overline{z}}\Big(\rho\frac{\partial \omega}{\partial
z}-\alpha_{1}\frac{\partial^{3}\omega}{\partial z^{2}
\partial\overline{z}})\Big\}-\mu\frac{\partial^{2}\omega}{\partial
z\partial\overline{z}}-\\&&\nonumber-2\beta_{3}\Big\{\frac{\partial^{2}\psi}
{\partial\overline{z}^{2}}\frac{\partial^{2}M}{\partial\overline{z}^{2}}
+\frac{\partial^{2}\psi} {\partial
z^{2}}\frac{\partial^{2}M}{\partial
z^{2}}\Big\}-\\&&-\beta_{3}\Big\{\frac{\partial^{2}\omega}{\partial
z\partial\overline{z}}M + \frac{\partial\omega}{\partial
z}\frac{\partial
M}{\partial\overline{z}}+\frac{\partial\omega}{\partial\overline{z}
}\frac{\partial M}{\partial z}\Big\}=0 \eea where
\begin{equation}
\omega=-4\frac{\partial^{2}\psi}{\partial z\partial\overline{z}}
\end{equation}
\begin{equation}
M=32\frac{\partial^{2}\psi}
{\partial\overline{z}^{2}}\frac{\partial^{2}\psi} {\partial
z^{2}}\nonumber
\end{equation}

\section{Solutions}
On integrating equation (12), we obtain the general form of the
exact solutions
\begin{equation}
\psi=-\frac{1}{4}\int\int\omega\, dz\,d\overline{z}+A+\overline{A}
\end{equation}
where $A \,\,\,\mbox{and}\,\,\,\overline{A}$  are complex functions
and $\overline{A}$  is the complex conjugate of $A$.

To determine the solution of equation (11), our strategy will be to
specify the vorticity function $\omega$  and determine the
condition, which the functions $A
\,\,\,\mbox{and}\,\,\,\overline{A}$  must satisfy. The
condition is obtained from equation (11) utilizing equation (13).\\

I. \,\,\, When vorticity $\omega$  is constant say $\omega_{\circ}$
, then equation (13) yields
\begin{equation}
\psi=-\frac{1}{4}\omega_{\circ}z\overline{z}+A+\overline{A}
\end{equation}
Equation (11), on utilizing equation (14), yields
\begin{equation}
\beta_{3}\Big(\frac{\partial^{4}A}{\partial
z^{4}}+\frac{\partial^{4}\overline{A}}{\partial\overline{z}^{4}}\Big)=0,\,\,\,\,\,
\frac{\partial^{2}A}{\partial z^{2}}\neq
0,\,\,\,\,\,\frac{\partial^{2}\overline{A}}{\partial\overline{z}^{2}}\neq
0.\nonumber
\end{equation}
When $\beta_{3}\neq 0$,\,\,\,then
\begin{equation}
A=\frac{\iota
a_{1}z^{4}}{24}+\frac{a_{2}z^{3}}{6}+\frac{a_{3}z^{2}}{2}+a_{4}z+a_{5}
\end{equation}
where $a_{i}$'s  are all complex constants. The velocity components
$u$ and $v$ are given by \bea\nonumber&& u=\frac{\iota
\omega_{\circ}}{4}(z-\overline{z})-\iota\Big\{-\frac{\iota
\overline{a_{1}}\overline{z^{3}}}{6}+\frac{\overline{a_{2}}\overline{z^{2}}}{3}
+\overline{a_{3}}\overline{z}+\overline{a_{4}}-\\&&-\frac{\iota
a_{1}z^{3}}{6}-\frac{a_{2}z^{2}}{3}-a_{3}z-a_{4}\Big\};\nonumber
\eea \bea\nonumber&&
v=\frac{\omega_{\circ}}{4}(z+\overline{z})-\Big\{\frac{\iota
a_{1}z^{3}}{6}+\frac{a_{2}z^{2}}{3}+a_{3}z+a_{4}-\\&&-\frac{\iota
\overline{a_{1}}\overline{z^{3}}}{6}+\frac{\overline{a_{2}}\overline{z^{2}}}{3}
+\overline{a_{3}}\overline{z}+\overline{a_{4}}\Big\}\nonumber \eea
Equation (14), on utilizing equation (15), yields \bea \nonumber &&
\psi=-\frac{\omega_{\circ}z\overline{z}}{4} +\frac{\iota
(a_{1}z^{4}-\overline{a_{1}z^{4}})}{24}+\frac{(a_{2}z^{3}+\overline{a_{2}z^{3}})}{6}
+\\&& +\frac{(a_{3}z^{2}+\overline{a_{3}z^{2}})}{2}+
(a_{4}z+\overline{a_{4}z})+a \eea where $a = a_5 +
\overline{a_{5}}$\,\,.\\
The stream function $\psi$ in equation (16) is represented
graphically in Fig.1 with $\omega_0 = -1 ,\,\, a_1 = 1 + 2\iota
,\,\, a_2 = 1 + \iota ,\,\, a_3 = 1 + 5\iota ,\,\, a_4 = 2 +
0.5\iota ,\,\, a = 2$ and $ x\, , y \in [-1,\,1]$.
\begin{center}
\epsfig{file=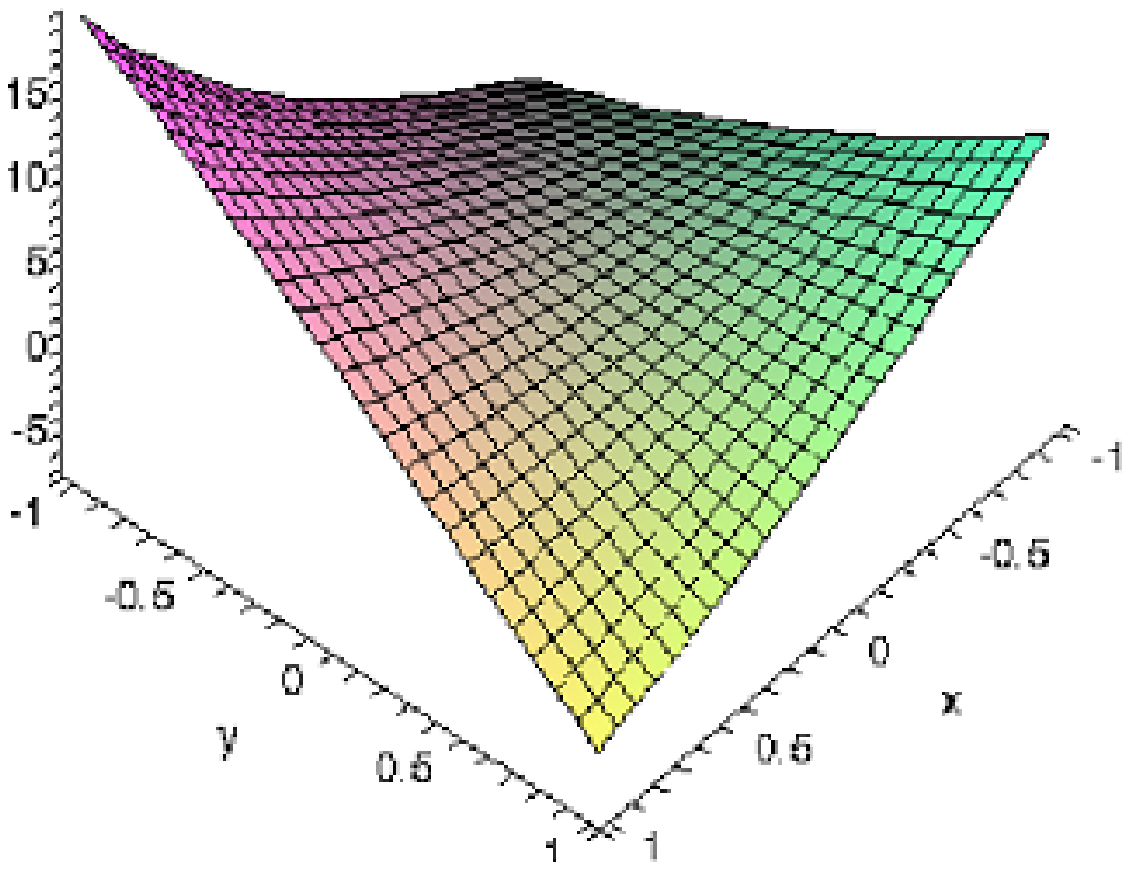,width=6.0 cm}\\Figure No. 1: Graphical
representation of equation (16)
\end{center}
When $\beta_{3}$  = 0, then equation (15) is identically satisfied
and the complex $A$ becomes arbitrary, due to which $A$  enables us
to construct a large number of streamfunction $\psi$  and hence a
large number of solutions to the flow equations. We mention that by
taking $\omega = 0$  or appropriately choosing complex constants
$a_{3},\,\,a_{4},\,\,a_{5},\,\,\mbox{and}\,\,a_{1}=0=a_{2}$ in
equation (15), or by taking $A=\iota a\log z$ (for$\,\,\beta_{3}$ =
0), we get all the solutions of Moro et al. \cite{Moro}.

Further more, if we take the function $A=(c_{1}+\iota c_{2})z^{2}$
or\, $A=(c_{1}+\iota c_{2})\ln z$ and choose appropriately the
constants or apply the appropriate boundary conditions we get the
plane Couette flow, the flow due to a spiral vortex at the origin
and the flows having streamlines as a family of ellipses, concentric
circles, rectangular hyperbolae.\\\\
II.\,\,\,\,\, When $\omega$  is non-constant, the solutions of
equation (11) are determined as follows: \\\\
(i) \,\,\, When $\omega=m_{1}z+\overline{m_{1}}\overline{z}$  ,  the
equation (11) yields \bea\nonumber && Im\Big\{-\frac{m^{2}_1
z^{2}}{8}+m_{1}\frac{\partial \overline{A}}{\partial
\overline{z}}\Big\}=\lambda\Big\{6m_{1}\overline{m_{1}}(m_{1}z
+\overline{m_{1}}\overline{z})-\\&&\nonumber
-8\Big(m^{2}_1\frac{\partial^{2}\overline{A}}{\partial
\overline{z}^{2}}+\overline{m^{2}_1}\frac{\partial^{2}A}{\partial
z^{2}}\Big)
-16\Big(m^{2}_1\overline{z}\frac{\partial^{3}\overline{A}}{\partial
\overline{z}^{3}}+\overline{m^{2}_1}z\frac{\partial^{3}A}{\partial
z^{3}}\Big)\\&&\nonumber+64
\Big(m_{1}\frac{\partial^{3}\overline{A}}{\partial\overline{z}^{3}}
\frac{\partial^{2}A}{\partial
z^{2}}+\overline{m_{1}}\frac{\partial^{3}A}{\partial z^{3}}
\frac{\partial^{2}\overline{A}}{\partial\overline{z}^{2}}\Big)
+\\&&+64\frac{\partial^{3}A}{\partial z^{3}}
\frac{\partial^{3}\overline{A}}{\partial\overline{z}^{3}}(m_{1}z
+\overline{m_{1}}\overline{z})\Big\} \eea where
$\lambda=\frac{\beta_{3}}{\rho}$ , and $m_{1}$ is the complex
constant.\\ The L.H.S of equation (17) suggests to assume
\begin{equation}
\frac{\partial\overline{A}}{\partial\overline{z}}=\lambda_{1}\overline{z}^{2}
+\lambda_{2}\overline{z}+\lambda_{3}\nonumber
\end{equation}
This on putting in equation (17), gives
\begin{equation}
\lambda_{1}=-\frac{\overline{m^{2}_1}}{8m_{1}},\,\,\,\,\,\,
\,\,\,\,\lambda_{2}=40\iota\lambda \overline{m^{2}_1},
\,\,\,\,\,\,\,\,\,\,\lambda_{3}=0.\nonumber
\end{equation}
The solution of equation (17), therefore, is
\begin{equation}
A=-\frac{m^{2}_1 z^{3}}{24\overline{m_{1}}}-20\iota\lambda
m^{2}_1z^{2}+m_{2}
\end{equation}
where $m_{2}$  is a complex constant.\\Equation (13), using (18)
becomes \bea\nonumber&&
\psi=-\frac{z\overline{z}}{8}(m_{1}z+\overline{m_{1}}\overline{z})-\frac{\Big(m^{3}_1z^{3}
+\overline{m^{3}_1}\overline{z^{3}}\Big)}{24m_{1}\overline{m_{1}}}+\\&&+20\iota\lambda\Big
(\overline{m^{2}_1}\overline{z^{2}}-m^{2}_1z^{2}\Big)+m \eea where
$m = m_2 + \overline{m_2}$\, .\\ The stream function $\psi$ in
equation (19) is represented graphically in Fig.2 with $m_1 = 1 +
2\iota\, ,\,\, m = 1\, ,\,\, \lambda = 0.3 $\,\, and \, $x ,\, y \in
[-1 ,\,1]$.
\begin{center}
\epsfig{file=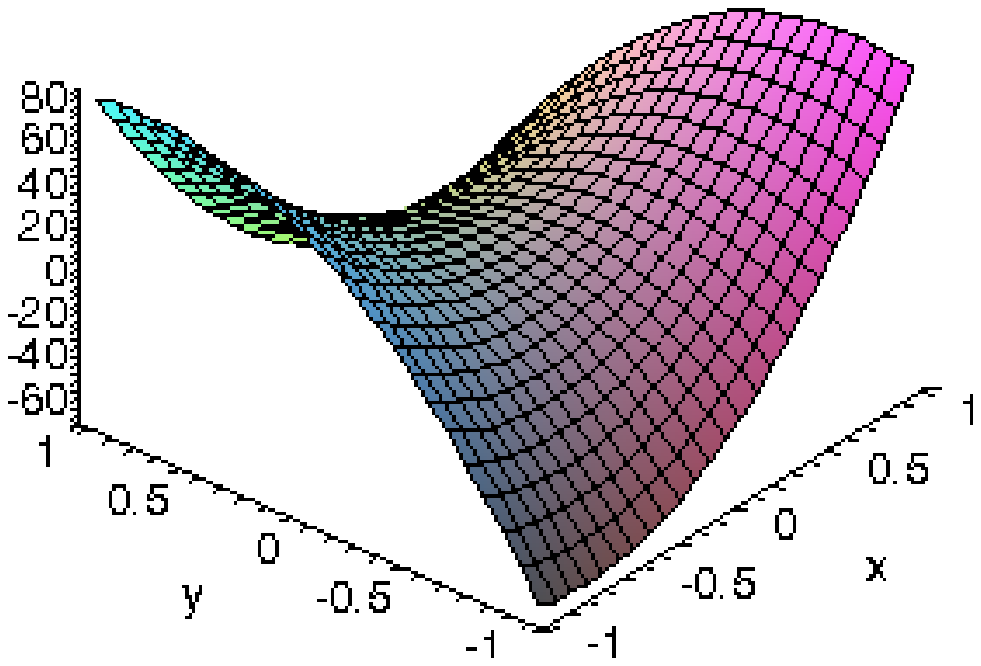,width=7.0 cm}\\Figure No. 2: Graphical
representation of equation (19).
\end{center}
(ii)\,\,\,  For $\omega=B(z+\overline{z})$ , the constant $B$  being
real, the equation (11) becomes \bea\nonumber &&
Im\Big\{-\frac{Bz^2}{8}+\frac{\partial\overline{A}}{\partial\overline{z}}\Big\}
=\lambda\Big\{6B^{2}(z+\overline{z})-
\\&&\nonumber-8B\Big(\frac{\partial^{2}A}{\partial
z^{2}}+\frac{\partial^{2}\overline{A}}{\partial\overline{z}^{2}}\Big)
-16B\Big(\overline{z}\frac{\partial^{3}\overline{A}}{\partial
\overline{z}^{3}}+z\frac{\partial^{3}A}{\partial z^{3}}\Big)+
\\&&\nonumber+64\Big(\frac{\partial^{2}A}{\partial
z^{2}}\frac{\partial^{3}\overline{A}}{\partial\overline{z}^{3}}
+\frac{\partial^{2}\overline{A}}{\partial\overline{z}^{2}}\frac{\partial^{3}A}{\partial
z^{3}} \Big)+\\&&\nonumber+64\frac{\partial^{3}A}{\partial z^{3}}
\frac{\partial^{3}\overline{A}}{\partial\overline{z}^{3}}(z+\overline{z})-4B\Big(z^{2}
\frac{\partial^{4}A}{\partial
z^{4}}+\overline{z^{2}}\frac{\partial^{4}\overline{A}}{\partial\overline{z}^{4}}\Big)
+\\&&+32\Big(z\frac{\partial^{2}\overline{A}}{\partial\overline{z}^{2}}
\frac{\partial^{4}A}{\partial
z^{4}}+\overline{z}\frac{\partial^{2}A}{\partial
z^{2}}\frac{\partial^{4}\overline{A}}{\partial(\overline{z})^{4}}\Big)-\\&&\nonumber-
\frac{64}{B}\Big(\frac{\partial^{2}A^{2}}{\partial
z^{2}}\frac{\partial^{4}\overline{A}}{\partial\overline{z}^{4}}
+\frac{\partial^{2}\overline{A^{2}}}{\partial\overline{z}^{2}}\frac{\partial^{4}A}{\partial
z^{4}}\Big)\Big\}\eea The L.H.S of equation (20) suggests
$\frac{\partial \overline{A}}{\partial\overline{z}}$ to be a
polynomial in $\overline{z}$ of degree two and therefore on
substituting
\begin{equation}
\frac{\partial\overline{A}}{\partial\overline{z}}=l_{1}\overline{z}^{2}
+l_{2}\overline{z}+\lambda_{4},\nonumber
\end{equation}
in equation (20), we get
\begin{equation}
l_{1}=-\frac{B}{8},\,\,\,\,\,\,\,\,\,\,\,\,\,\,\,\,\,\,\,\,\,\,l_{2}=40\iota\lambda
B^{2},\,\,\,\,\,\,\,\,\,\,\,\,\,\,\,\,\,\,\,\,\,\,\,\,\,\,\,\,\,\,\lambda_{4}=0\nonumber
\end{equation}
Hence
\begin{equation}
\overline{A}=-\frac{B\overline{z^{3}}}{24}+20\iota\lambda
B^{2}\overline{z}+m_{3}
\end{equation}
where $m_{3}$  is an arbitrary complex constant.\\ Using equation
(21), equation (13) implies,
\begin{equation}
\psi=-\frac{Bz\overline{z}}{8}(z+\overline{z})-\frac{B}{24}\Big(z^{3}
+\overline{z^{3}}\Big)+20\iota B^{2}\lambda(\overline{z}-z)+n
\end{equation}
where $n = m_3 + \overline{m_3}$\,\,.\\ The stream function $\psi$
in equation (22) is represented graphically in Fig.3 with $B = -2\,
, \,\, \lambda = 2\, , \,\, n = 1$ \,\, and \,\, $x \in [-10 ,\,
10]\, ,\,\, y \in [0 ,\, 10]$\,.
\begin{center}
\epsfig{file=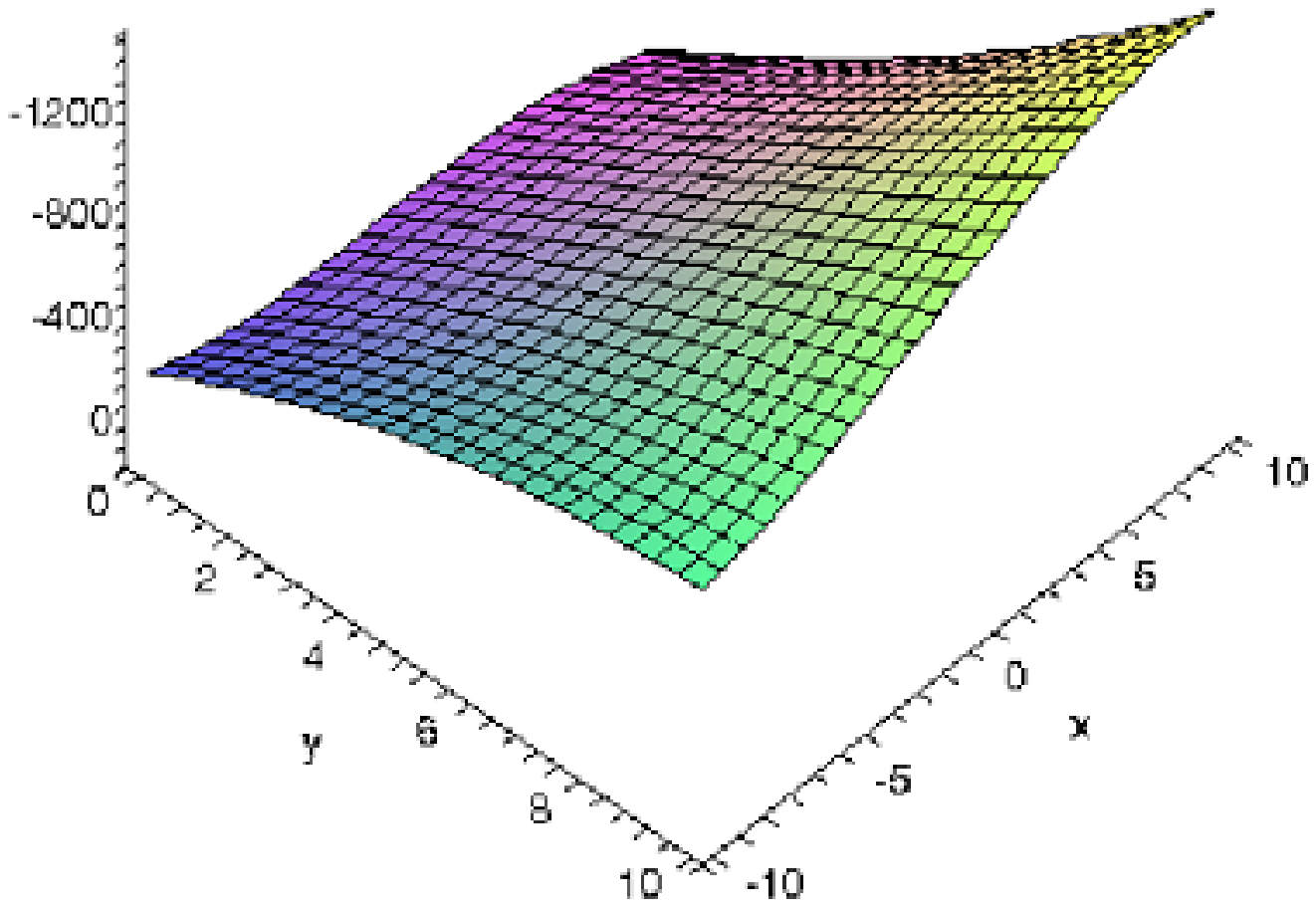,width=6.0 cm}\\Figure No. 3: Graphical
representation of equation (22).
\end{center}

( iii ) When $\omega=D(z+\overline{z}+E),$ $D$\,and\,$E$  being
real, then equation (11) yields
 \bea\nonumber &&
Im\Big\{-\frac{Dz^2}{8}-\frac{DE
z}{4}+\frac{\partial\overline{A}}{\partial\overline{z}}\Big\}
=\lambda\Big\{6D^{2}(z+\overline{z})-\\&&\nonumber
-8D\Big(\frac{\partial^{2}A}{\partial
z^{2}}+\frac{\partial^{2}\overline{A}}{\partial\overline{z}^{2}}\Big)-16D\Big(z\frac{\partial^{3}A}{\partial
z^{3}}+\overline{z}\frac{\partial^{3}\overline{A}}{\partial
\overline{z}^{3}}\Big)+\\
&&\nonumber+64\Big(\frac{\partial^{2}A}{\partial
z^{2}}\frac{\partial^{3}\overline{A}}{\partial\overline{z}^{3}}
+\frac{\partial^{2}\overline{A}}{\partial\overline{z}^{2}}\frac{\partial^{3}A}{\partial
z^{3}} \Big)+64\Big(z\frac{\partial^{3}A}{\partial z^{3}}
\frac{\partial^{3}\overline{A}}{\partial\overline{z}^{3}}+
\\&&\nonumber+\overline{z}\frac{\partial^{3}A}{\partial z^{3}}
\frac{\partial^{3}\overline{A}}{\partial\overline{z}^{3}}\Big)
+4D^{2}E+64E\frac{\partial^{3}A}{\partial z^{3}}
\frac{\partial^{3}\overline{A}}{\partial\overline{z}^{3}}-\\&&\nonumber
-4D\Big(\overline{z^{2}}
\frac{\partial^{4}\overline{A}}{\partial\overline{z}^{4}}+
z^{2}\frac{\partial^{4}A}{\partial z^{4}}\Big)
-\frac{64}{D}\Big(\frac{\partial^{2}A^{2}}{\partial
z^{2}}\frac{\partial^{4}\overline{A}}{\partial\overline{z}^{4}}
+\\&&\nonumber+\frac{\partial^{2}\overline{A^{2}}}{\partial\overline{z}^{2}}\frac{\partial^{4}A}{\partial
z^{4}}\Big)+32\Big(\frac{\partial^{2}\overline{A^{2}}}{\partial\overline{z}^{2}}
\frac{\partial^{4}A}{\partial
z^{4}}+\frac{\partial^{2}A^{2}}{\partial
z^{2}}\frac{\partial^{4}\overline{A}}{\partial\overline{z}^{4}}\Big)\Big\}\nonumber
\eea Following the same procedure as that of previous case we find
\begin{equation}
\overline{A}=-\frac{D\overline{z^{3}}}{24}-\frac{DE\overline{z^{2}}}{8}+20\iota\lambda
D^{2}\overline{z^{2}}+20\iota\lambda D^{2}E \overline{z}+m_{4}
\end{equation}
where $m_{4}$  is an arbitrary complex constant.\\Using (23),
equation (13) becomes \bea\nonumber&&
\psi=-\frac{Dz\overline{z}}{8}(z+\overline{z}+2E)-\frac{D}{24}\Big(z^{3}+\overline{z^{3}}\Big)
-\frac{DE} {8}\Big(z^{2}+\\&&+\overline{z^{2}}\Big)+20\iota\lambda
D^{2}\Big(\overline{z^{2}}-z^{2}\Big)+20\iota\lambda
D^{2}E(\overline{z}-z)+ q \eea where $ q = m_4 +
\overline{m_4}$\,.\\The stream function $\psi$ in equation (24) is
represented graphically in Fig.4 with $ D = E = 1\, ,\,\, \lambda =
2\, ,\,\, q = -1 $\,\, and \, $ x ,\, y \in [-10 ,\, 10]$\,.
\begin{center}
\epsfig{file=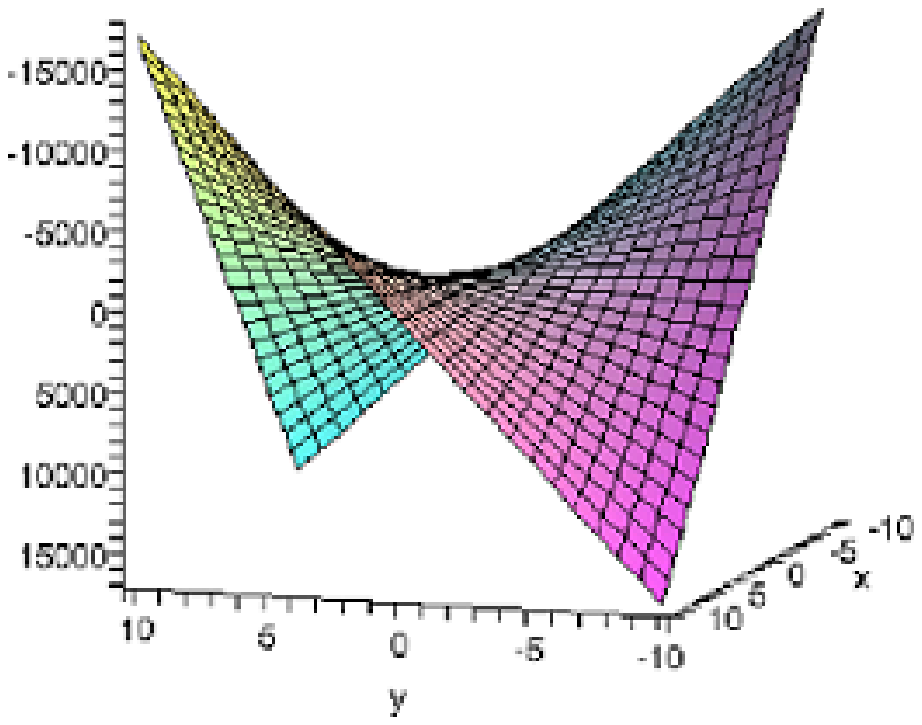,width=7.0 cm}\\Figure No. 4: Graphical
representation of equation (24).
\end{center}
(iv)   When $\omega=B\iota(z-\overline{z})$, then equation (11) is
satisfied provided
\begin{equation}
\overline{A}=-\frac{B\iota \overline{z^{3}}}{24}-20\iota\lambda
B^{2}\overline{z^{3}}+m_{5}\nonumber
\end{equation}
where $m_{5}$ is an arbitrary complex constant and the
streamfunction \,$\psi$\, is given by
\begin{equation}
\psi=-\frac{B\iota}{8}z\overline{z}\Big(z-\overline{z}\Big)+\frac{B\iota}{24}\Big(z^{3}
-\overline{z^{3}}\Big) +20\iota\lambda
B^{2}\Big(z^{3}-\overline{z^{3}}\Big)+ r
\end{equation}
where $ r = m_5 + \overline{m_5}$\,.\\ The stream function $\psi$ in
equation (25) is represented graphically in Fig.5 with $ B = -5\,
,\,\, \lambda = 3\, ,\,\, r = 10 $ \,\, and \,\, $ x \in [-2 ,\,
10]\, ,\,\, y \in [0 ,\, 8]$\,.
\begin{center}
\epsfig{file=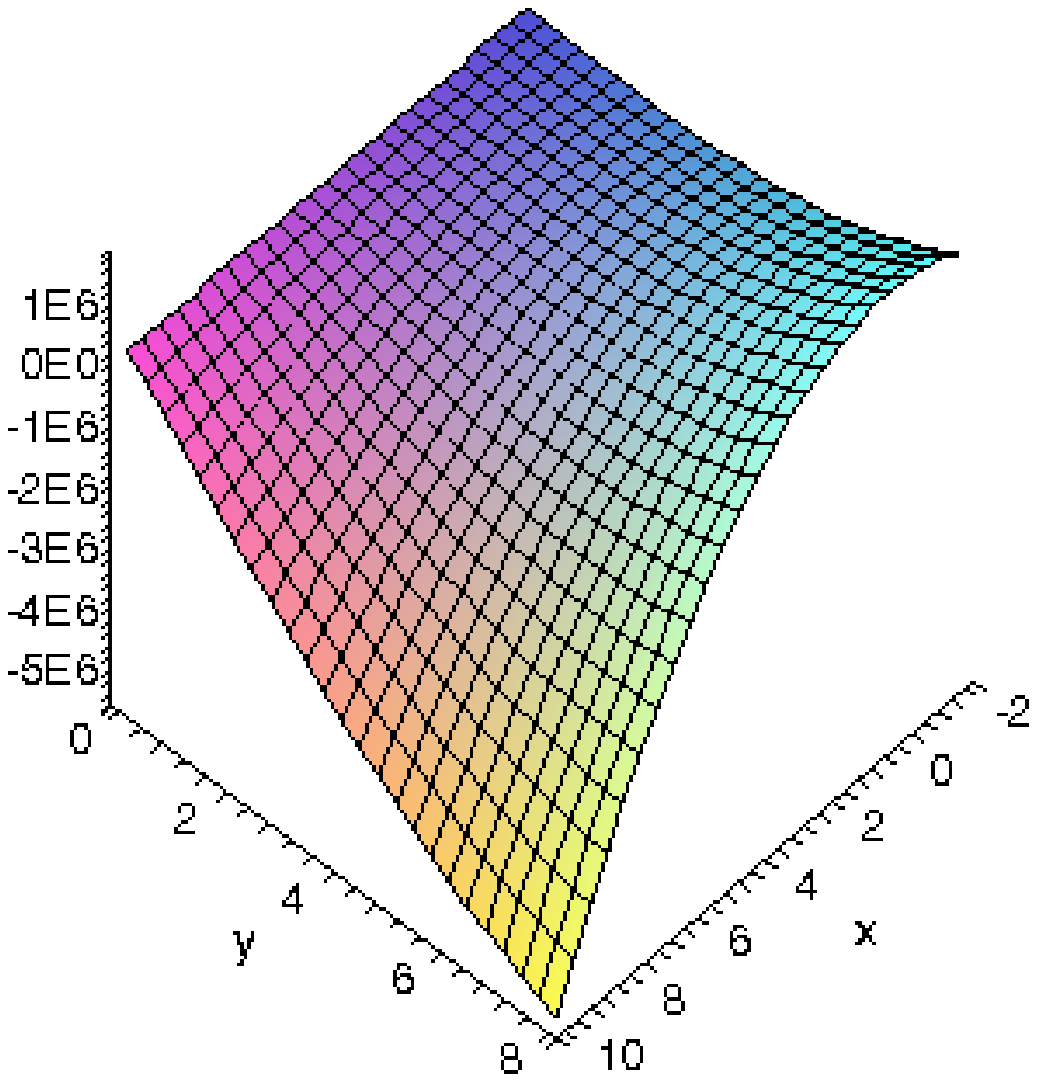,width=6.0 cm}\\Figure No. 5: Graphical
representation of equation (25).
\end{center}
( v )  For $\omega=Bln(z\overline{z})+D_{1}$ , the equation (11) is
satisfied provided
\begin{equation}
\overline{A}=m_{6}ln\overline{z}+m_{7}
\end{equation}
where $ m_{6},\, B,\, D_{1},$ are real constants and $m_{7}$   is
complex arbitrary constant.\\According to equation (26), the
streamfunction $\psi$  becomes
\begin{equation}
\psi=-\frac{B}{4}z\overline{z}\Big\{ln(z\overline{z})-2\Big\}
-\frac{D_{1}}{4}z\overline{z}+m_{6}ln(z\overline{z}) + s
\end{equation}
where $ s = m_{7} + \overline{m_{7}}$\,.\\ The stream function
$\psi$ in equation (27) is represented graphically in Fig.6 with $
m_6 = 2\, ,\,\,\, s = 4 $ \,\, and \,\, $ x \in [-10 ,\, 2]\, ,\,\,
y \in [-1 ,\, 2]$\,.
\begin{center}
\epsfig{file=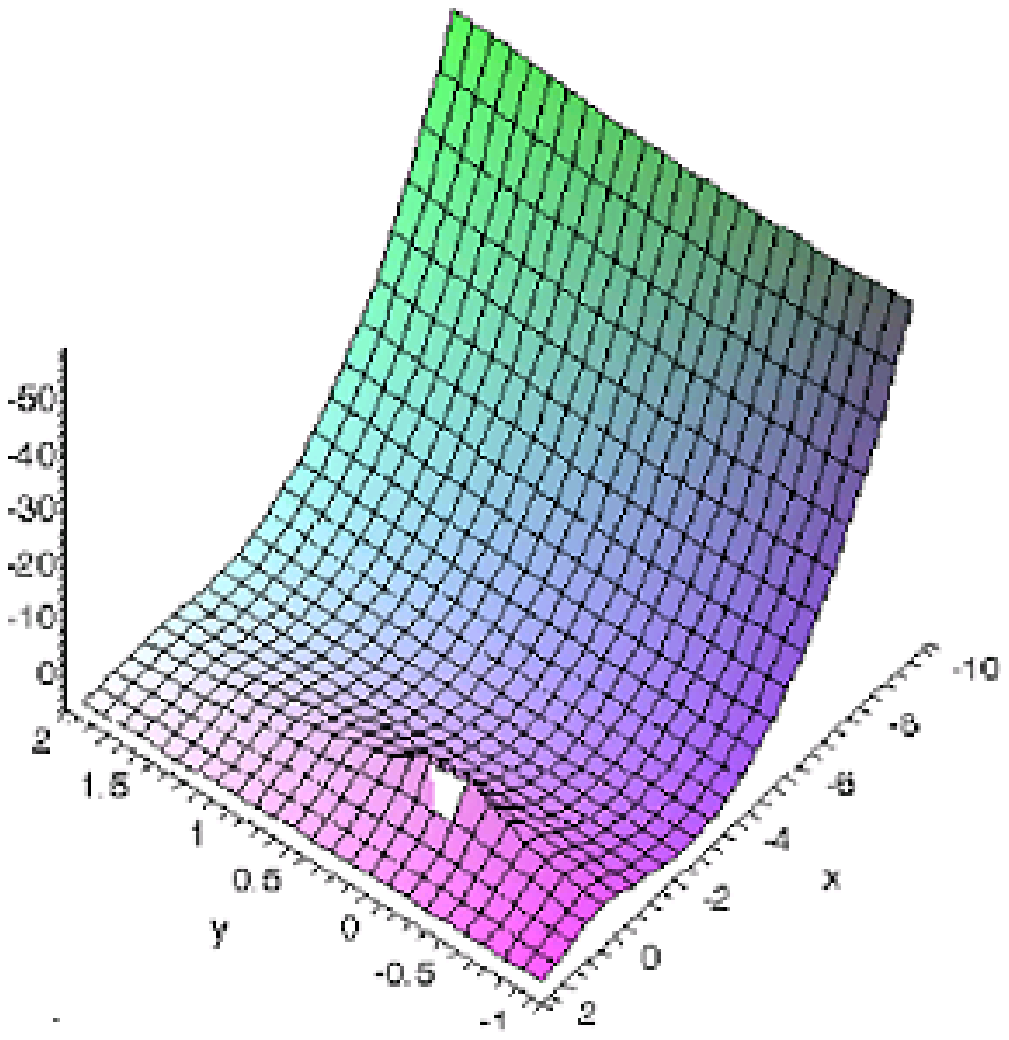,width=6.0 cm}\\Figure No. 6: Graphical
representation of equation (27).
\end{center}
The stream function $\psi$ in equation (27) represents the motion
whose streamlines are concentric circles with constant speed along
each streamline for the appropriate choice of the constants.\\

( vi )   When $\omega=Bz\overline{z},\,\,B$\, being real constant,
then equation (11) is satisfied provided
\begin{equation}
\overline{A}=\frac{\iota\mu}{\rho}ln\overline{z}+ m_{8}
\end{equation}
where $m_{8}$  is a complex constant.\\
According to equation (28), the stream function $\psi$  becomes
\begin{equation}
\psi=-\frac{B}{16}(z\overline{z})^{2}-\frac{\iota\mu}{\rho}ln\Big(\frac{z}{\overline{z}}\Big)
+ t
\end{equation}
where $ t = m_{8} + \overline{m_{8}}$\,.\\The stream function $\psi$
in equation (29) is represented graphically in Fig.7 with $ B = 1\,
,\,\, \mu = 12\, ,\,\, \rho = 1\, ,\,\, t = 2$ \,\, and \,\, $ x \,
,\,\, y \in [-1 ,\, 10]$\,.
\begin{center}
\epsfig{file=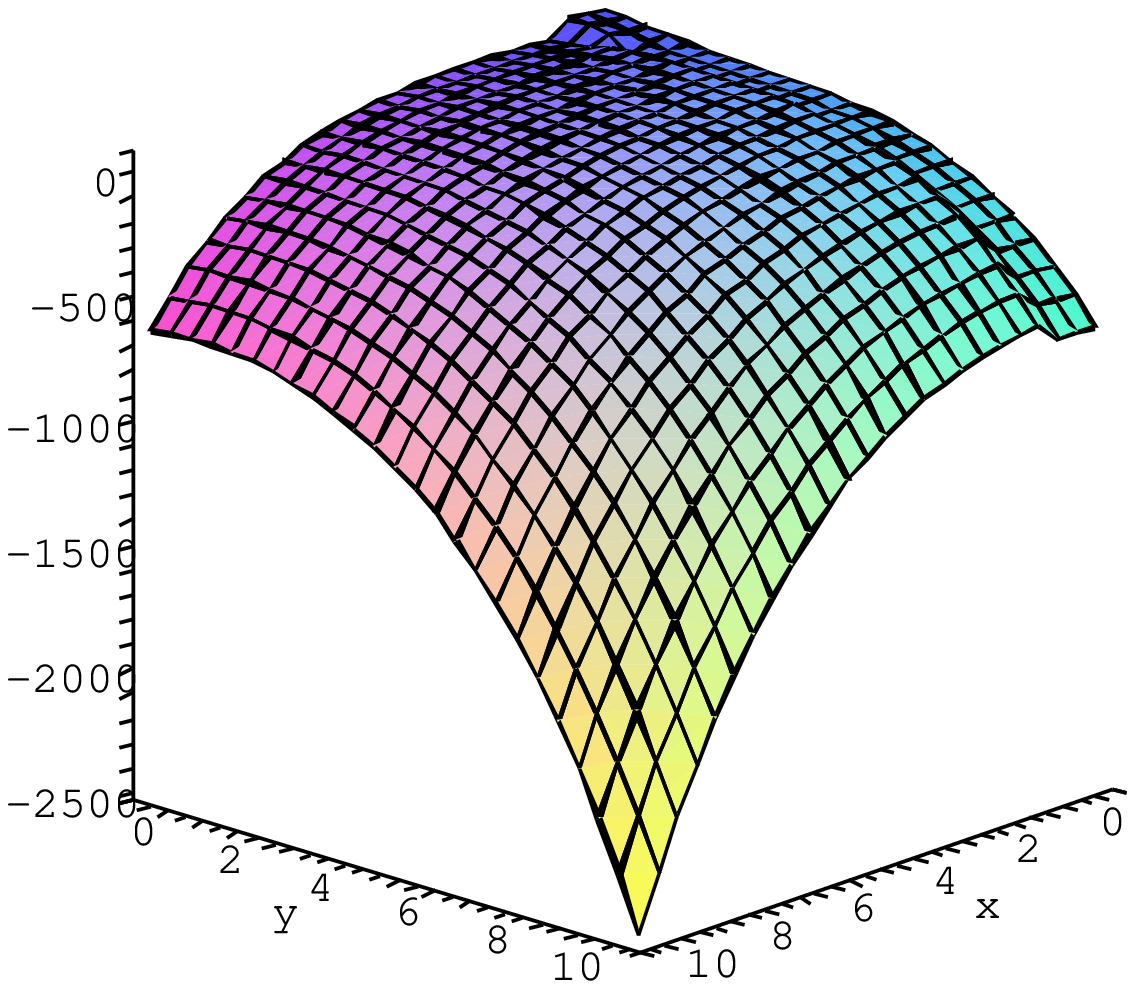,width=7.0 cm}\\Figure No. 7: Graphical
representation of equation (29).
\end{center}
\section{Conclusions}

In this paper, we reconsidered the flow equations of Moro et al.
\cite{Moro} with the objective of determining some exact solutions.

For this purpose the vorticity function $\omega$  and the stream
function $\psi$  are expressed in terms of complex variables and
complex function. The condition which the complex functions must
satisfy is determined through the equations for the generalized
energy function $h$  by using the compatibility condition
$\frac{\partial^2 h}{\partial x
\partial y}= \frac{\partial^2 h}{\partial y \partial x}$.

Some exact solutions to the flow equations are determined using the
condition for the complex functions. The solutions presented in this
paper admit, as particular cases, all the solutions of Moro et al.
\cite{Moro} by appropriately choosing the complex functions or the
arbitrary constants therein.

\begin{acknowledgments}
I wish to thank the reviewers for their valuable comments and
suggestions, which significantly improved the paper.

I am also thankful to Prof. C. Fetecau (Romania) and Prof. D. Vieru
(Romania) for productive scientific discussions and valuable
suggestions.
\end{acknowledgments}

\bibliography{apssamp}

\end{document}